\documentclass[smallcondensed]{svjour3}    
\smartqed  
\usepackage{graphicx}
\usepackage{fix-cm}
\usepackage{amssymb}
\usepackage{amsmath}
\def\pd#1#2{\frac{\partial #1}{\partial#2}}
\def\matriz#1#2{\left( \begin{array}{#1} #2 \end{array}\right) }
\journalname{Int J Theor Phys}
\begin{document}

\title{A geometric approach to integrability of 
Abel differential equations
}


\author{Jos\'e F. Cari\~nena \and Javier de Lucas \and Manuel F. Ra\~nada
}


\institute{Jos\'e F. Cari\~nena \at
Departamento de F\'{\i}sica Te\'orica and IUMA,  Universidad de Zaragoza, 50009 Zaragoza, Spain \\
\email{jfc@unizar.es}           
        \and
Javier de Lucas \at Institute of Mathematics, Polish Academy of Sciences,  ul. Sniadeckich 8, PO Box 21, 
00-956 Warszawa, Poland
\email{delucas@impan.gov.pl}
   \and
Manuel F. Ra\~nada \at
Departamento de F\'{\i}sica Te\'orica and IUMA,  Universidad de Zaragoza, 50009 Zaragoza, Spain \\
\email{mfran@unizar.es}           
}

\date{Received: date / Accepted: date}

\maketitle

\begin{abstract}
A geometric approach is used to study the Abel first order differential equation of the first kind. The
approach is based on the recently developed theory of quasi-Lie systems which 
allows us to characterise some particular examples of integrable Abel equations.
Second order Abel equations will be discussed and the
inverse problem of the Lagrangian
dynamics is analysed:  the existence of two alternative
Lagrangian formulations  is proved, both Lagrangians being of a non-natural
class. The study is carried out 
by means of the Darboux polynomials
and Jacobi multipliers.

\end{abstract}

\section{Introduction: Abel differential equations}
\label{intro}

Nonlinear differential equations play a relevant role in the study of evolution in terms of an 
evolution parameter, what has motivated physicists' interest during the last forty years. The 
lack of a general procedure for determining their general solutions justifies the analysis of 
particular instances where solutions can be investigated via algebraic or geometric methods. An 
important example are Lie systems, which admit a superposition rule giving the general solution 
in terms of a finite set of particular solutions. These systems describe the integral curves of 
$t$-dependent vector fields which are $t$-dependent linear combinations of vector fields closing
 on a finite-dimensional real Lie algebra (see e.g. \cite{CGM00} and \cite{CGM07} for a modern 
  presentation of the theory of Lie systems \cite{LS,PW}).

An interesting instance of Lie system is the Riccati equation, a generalisation of the 
inhomogeneous linear equation. One of its main characteristics is the existence of an action of the group of
curves in $SL(2,\mathbb{R})$ on the set of Riccati equations, what can be used for 
reduction of a given Riccati equation into a simpler one. This is for instance the case when a
 particular solution of a given equation is known \cite{CarRam}. Moreover, this property can be
  shown to be common to all Lie systems \cite{CGR99}.
 
A generalisation of Lie systems has recently been proposed \cite{CGL09,{CGL10}} allowing us to 
use some of the techniques applicable to Lie systems to this more general class of systems. This
 paper deals with a specific example, the Abel equation, with applications in many different
  branches of physics.

There are several differential equations called Abel equation. 
The first-order Abel equation of the first kind 
\begin{equation}
\dot x=A_0(t)+ A_1(t) x + A_2(t) x^2 + A_3(t) x^3 ,\label{firstkind}
\end{equation}
is a generalisation of the Riccati equation (the particular case $A_3=0$) introduced by Abel within the theory of elliptic functions, while the first-order Abel equation of the second kind is of the form \cite{AM,{LB}}:
\begin{equation}
(y+f(t))\, \dot y=B_0(t)+ B_1(t) y + B_2(t) y^2 + B_3(t) y^3,\label{secondkind}
\end{equation}
which reduces into one of the first kind by means of the transformation 
\begin{equation}
x=(y+f(t))^{-1}, \label{firsttosecond}
\end{equation}
with coefficients
$$
\begin{array}{rl}&A_0=-B_3,\quad A_1=3B_3\,f-B_2,\\
&A_2=-f'-3B_3\,f^2+2f\,B_2-B_1,\\
& A_3=f^3B_3-f^2\,B_2+f\,B_1-B_0.\end{array}
$$
Therefore, we can restrict ourselves to study the Abel equations of the first kind.

We can also consider the generalised first-order Abel equation \cite{Mo03}
\begin{equation}
\dot x=\sum_{k=0}^n A_k(t)\, x^k, \quad n>3,\quad n\in\mathbb{N}, \label{generalised}
\end{equation}
or the second-order Abel equation \cite{CGR09}:
\begin{equation}
 \frac{d^2 x}{dt^2} + 4  x^2 \Bigl(\frac{d x}{dt}\Bigr) +  x^5  =
0  \,.\label{AbelEq2}
\end{equation}
The $n$-order Abel equation is a linear combination of the different members of a hierarchy of 
$j$-order Abel equations with functions $p_j=p_j(t)$ as coefficients \cite{CGR09}:
\begin{equation}
 (p_0 {\mathbb{D}}_A^n + p_1 {\mathbb{D}}_A^{n-1} + \dots + p_{n-1} {\mathbb{D}}_A +  p_n)x + p_{n+1} = 0 \,.
\end{equation}
Here, ${\mathbb{D}}_A$ denotes the differential operator ${\mathbb{D}}_A={d}/{dt} + x^2$, and 
the action of ${\mathbb{D}}_A$ leads to a family of differential equations whose first members
 are given by a sequence ${\mathbb{D}}_A^ jx=0$, $j=0,1,2,\ldots$.

Abel equations appear in the reduction  of order of many second- and higher-order equations, and 
hence are frequently found in the modeling of real problems in many areas, for instance, Emden 
equation, $y''+(2/x)y'+y^n=0$,  Emden--Fowler  equation, $ y''=A\, \, x^n\, y^m\, (y')^l$,
the Van der Pol equation $y''-\epsilon(1-y^2)y'+\alpha\, y=0$, the Duffing equation, $y''+ay+b\, y^3=0$, and 
many other equations.
They also play a relevant role in the study of quadratic systems in the plane \cite{AGG}  and 
the center-focus problem \cite{BFY}.

The Li\'enard equation $x''+f(x)\,x'+g(x)=0$ was studied in 
 \cite{I08}. Defining $\xi(x)=x'$, it may be written
$\xi\,\xi'+f(x)\,\xi+g(x)=0$, which is an Abel equation of the second kind, related by $u=1/y$ (see (\ref{firsttosecond})) to the Abel equation of the first kind:
$u'=f(u)\, u^2+g(u)\, u^3$. In particular, the one studied by Bougoffa \cite{LB},
$\xi\,\xi'-\xi=\Phi(x)$  reduces to $u'+u^2+\Phi(x)\, u^3=0$.
For instance, the autonomous dissipative Milne--Pinney equation can be transformed into a second kind
Abel equation:
$$
y\left( \frac {dy}{dt} + 1\right) = \frac 1{t^3},
$$
which is related to the first kind Abel equation
$\dot x=x^2-\frac 1{t^3}x^3$.

The Abel equation was used by  Majorana in the study of  Thomas--Fermi equation:
\begin{equation}
y''=\frac{y^{3/2}}{x^{1/2}}, 
\end{equation}
which is a particular case of Emden--Fowler equation, and it is also used in the study of cosmological models \cite{GIM}.
The general solution of the cosmological Einstein-Friedmann equations for the universe
filled with a scalar field for a given potential can be expressed via the general solution of the Abel equation 
of the first kind \cite{YuYu}. Another more  recent application can be found in \cite{ZT}.

\section{Lie systems and superposition rules}
\label{lssr}

There are (systems of $n$) first-order differential equations (to be called Lie systems) that admit a superposition rule, i.e. a function 
$\phi:\mathbb{R}^{n(m+1)}\to \mathbb{R}^{n}$ such that the general solution can be written as $x=\phi(x_{(1)}, \ldots, x_{(m)},k)$, the simplest example 
being the  inhomogeneous linear equation 
\begin{equation}
\dot x=c_0(t)+ c_1(t)\,x, 
\end{equation}
for which there is a superposition rule involving two generic particular solutions:
\begin{equation}
x(t)=\phi(x_1(t),x_2(t);k)=x_1(t)+k\,(x_2(t)-x_1(t)).
\end{equation}
Moreover, it admits a solution in terms of two  {quadratures}:
$$x(t)=\exp\left(\int^tc_1(t')\, dt'\right)\left[C-\int^{t}\exp\left(\int^{t'}c_1(t'')\, dt''\right)c_2(t')\,dt'\right]
$$
Another well known example is the Riccati equation, i.e. the following nonlinear differential equation:
\begin{equation}
\dot x=c_0(t)+ c_1(t)\,x+c_2(t)\, x^2\,.
\end{equation}
This type of equations  does not admit integration by quadratures in the general case.
Their solutions  are the integral curves of the $t$-dependent vector field 
$$
X_t=c_0(t)X_0+c_1(t)X_1+c_2(t)X_2,
$$
which is a linear combination with $t$-dependent coefficients of the vector fields 
$$
X_{0} =\partial_{x}\,,        \quad
X_{1} =x\,\partial_{x}\, ,    \quad
X_{2} = x^2\,\partial_{x}\,,  \label{sl2gen}
$$
that close on a three-dimensional real Lie  algebra,
with defining relations
$$
[X_{0},X_{1}] = X_{0} \,,      \quad 
[X_{0},X_{2}] = 2X_{1}\,,     \quad
[X_{1},X_{2}] = X_{2} \,,
$$
isomorphic to the ${\mathfrak{sl}}(2,{\mathbb{R}})$ Lie algebra. 
Therefore it is a Lie system admitting a  {superposition rule} which turns out to be \cite{CMN} 
\begin{equation}
\phi(x_1,x_2,x_3;k)=\frac {k\, x_1(x_3-x_2)+x_2(x_1-x_3)}{k\,(x_3-x_2)+(x_1-x_3)}\ .
\end{equation}
The vector fields $X_{0}$, $X_{1}$ and $X_{2}$ are a basis of the fundamental vector fields relative to the action of the group 
$SL(2,\mathbb{R})$ on $\overline{\mathbb{R}}=\mathbb{R}\cup\{\infty\}$:
$$
\Phi(A,x)=\left\{
\begin{array}{cl}
\dfrac{\alpha x+\beta}{\gamma x+\delta}\ ,&\,\, \mbox{if\ }\  x\notin\{-\delta/\gamma ,\infty\},\cr
{\alpha}/{\gamma}\ ,&\,\, \mbox{if\ }\  x=\infty,\cr
\infty\ ,&\,\, \mbox{if\ }\  x=-\delta/\gamma,\cr
\end{array}\right.\mbox{with}\ 
A=\matriz{cc}{{\alpha}&{\beta}\\{\gamma}&{\delta}}\in{SL(2,\mathbb{R})}.
$$
A Riccati equation can be seen as a curve in ${\mathbb{R}}^3$ and each element $A(t)$ of the group ${\mathcal{G}}$
of smooth $SL(2, {\mathbb{R}})$-valued curves, ${\rm Map}({\mathbb{R}},\,SL(2,{\mathbb{R}}))$, transforms every curve $x(t)$  in $\bar{\mathbb{R}}$ into a new one $\bar x(t)=\Phi({A}(t),x(t))$ 
satisfying a new Riccati equation with coefficients $\bar c_2, \bar c_1, \bar c_0$: 
$$
\begin{array}{rcl}
\bar c_2&=&{\delta}^2\,c_2-\delta\gamma\,c_1+{\gamma}^2\,c_0+\gamma{\dot{\delta}}-\delta \dot{\gamma}\ ,\nonumber \cr
\bar c_1&=&-2\,\beta\delta\,c_2+(\alpha\delta+\beta\gamma)\,c_1-2\,\alpha\gamma\,c_0   
       +\delta\dot{\alpha}-\alpha \dot{\delta}+\beta \dot{\gamma}-\gamma \dot{\beta}\ ,   \cr
\bar c_0&=&{\beta}^2\,c_2-\alpha\beta\,c_1+{\alpha}^2\,c_0+\alpha\dot{\beta}-\beta\dot{\alpha} \ .
 \nonumber 
\end{array}
$$
These expressions define an affine action of the group  ${\mathcal{G}}$ on the set of 
Riccati equations. An appropriate choice for the curve, $A(t)$,  transforms the original Riccati
equation into a simpler one, for instance an integrable by quadratures one.
The conditions for the existence of such a transformation provide us with integrability 
conditions \cite{CLR10}. Moreover, similar properties hold for any other Lie system.
\section{Geometric approach to the Abel equation of the first kind}

Each $t$-{dependent vector field} whose integral curves are the solutions of an Abel equation is
 a   linear combination with $t$-dependent coefficients of the vector fields in the linear space  
$V_{{\rm Abel}}({\mathbb{R}})$ defined by
$$
V_{{\rm Abel}}({\mathbb{R}})=\left\langle\partial_x,x\partial_x,x^2\partial_x,x^3\partial_x\right\rangle.
$$
Such a $t$-dependent vector field is not, in general, a Lie system. Indeed, consider a 
$t$-dependent vector field $X(t,x)=x^2\partial_x+h(t)x^3\partial_x$, with a 
non-constant function $h$. If $X(t,x)$ determines a Lie system, there must exist 
a finite-dimensional Lie algebra $V_0(\mathbb{R})$ including the vector fields 
$x^2\partial_x$ and $x^3\partial_x$ such that $X\in V_0(C^{\infty}(\mathbb{R}))$. 
Nevertheless, this is impossible as such a Lie algebra should contain the successive 
Lie brackets of $x^2\partial_x$ and $x^3\partial_x$, which span an infinite family of linearly independent vector fields (note that $[x^2\partial_x,x^n\partial_x]=(n-2)x^{n+1}\partial_x$). 

As pointed out in a recent paper \cite{CGL09}, we can however look for a linear subspace $W(\mathbb{R})\subset V_{{\rm Abel}}(\mathbb{R})$  such that $W(\mathbb{R})$  is a Lie algebra satisfying that
$[W(\mathbb{R}),V_{{\rm Abel}}(\mathbb{R})]\subset V_{{\rm Abel}}(\mathbb{R})$. In such a case   the flows of time-dependent vector fields with values in $W(\mathbb{R})$ leave invariant the 
linear space $V_{{\rm Abel}}(\mathbb{R})$ and starting with a given Abel equation we obtain a new Abel equation from each flow. Maybe, by an appropriate flow, the transformed Abel equation becomes a Lie system. In this case the initial Abel equations is called a quasi-Lie system.

Our aim is, given $Y=f(x)\,\partial_x\in\mathfrak{X}(\mathbb{R})$, to  determine $f$ such  that
 $[Y,V_{Abel}(\mathbb{R})]\subset V_{Abel}(\mathbb{R})$. In particular,
$$
\left[Y,{\partial_x}\right]=-f'(x)\partial_x\in V_{{\rm Abel}}(\mathbb{R}).
$$
Hence, $-f'(x)=c_3 x^3+c_2x^2+c_1x+c_0$ and $f(x)=-\frac{c_3}4 x^4-\frac{c_2}{3}x^3-\frac{c_1}{2}x^2+c_0x+c_{-1}$. Using now the condition 
$\left[Y,x\,{\partial_x}\right]=(f(x)-xf'(x))\partial_x\in V_{{\rm Abel}}(\mathbb{R}),$
we obtain that $c_3=0$, and $\left[Y,x^2{\partial_x}\right]=(2xf(x)-x^2f'(x))\partial_{x}\in V_{{\rm Abel}}(\mathbb{R})$
yields that  $c_2=0$.
Moreover, as $Y$ must satisfy that $\left[Y,x^3\, {\partial_x}\right]=(3x^2f(x)-x^3f'(x))\partial_{x}\in V_{{\rm Abel}}(\mathbb{R}),$
we also see that $c_1=0$.
Consequently, every symmetry group of $V_{{\rm Abel}}(\mathbb{R})$ admits a Lie algebra of fundamental vector fields $W_0$ contained in the Lie algebra
$$
W_{{\rm Abel}}=\left\langle\partial_{x},x\partial_{x}\right\rangle.
$$
We find in this way the so-called structure invariance group \cite{N08}, which turns out to be the affine group in one dimension.
 
Correspondingly, the set of first-order Abel equations of the first kind is  invariant under all the following transformations \cite{chebterr00}
 \begin{equation}
 \bar x(t)=\alpha(t)\,x(t)+\beta(t), \qquad \alpha(t)\ne 0,\label{stptr}
 \end{equation}
 Under such a transformation the given Abel equation (\ref{firstkind}) becomes a new Abel equation 
 $\dot {\bar x}=\bar A_0(t)+ \bar A_1(t)\,\bar  x + \bar A_2(t) \,\bar x^2 + \bar A_3(t)\,\bar  x^3$, with
\begin{equation}
\begin{aligned}
\bar A_3(t)&=A_3(t)\alpha^2(t),\\
\bar A_2(t)&=\alpha(t)(3A_3(t)\beta(t)+A_2(t)),\\
\bar A_1(t)&=3A_3(t)\beta^2(t)+2A_2(t)\beta(t)+A_1(t)-\dot\alpha(t)\alpha^{-1}(t),\\
\bar A_0(t)&=\alpha^{-1}(t)\left(A_3(t)\beta^3(t)+A_2(t)\beta^2(t)+A_1(t)\beta(t)+A_0(t)-\dot\beta(t)\right).\\
\end{aligned}\label{abeleqtrans}
\end{equation}
This is a very useful property: the  action of the group of curves in the affine group produces 
  orbits, i.e. equivalence classes of Abel equations, with the same properties of integrability
   or existence of superposition rules. For instance, Abel equations of an orbit containing an 
   integrable by quadratures Riccati equation are also integrable by quadratures. 
The orbits are characterized by different values of invariant functions.

One must determine the Lie algebras contained in $V_{\rm Abel}$ reachable by the set of transformations 
(the coefficient of $X_3$ cannot be transformed to be zero, see first equation in (\ref{abeleqtrans})).
It is possible to check that the only subalgebra of dimension  three is the one of Riccati equation which is not reachable by the considered set of transformations. On the other hand, there is a one-parameter family of two-dimensional subalgebras of $V_{\rm Abel}$ reachable from a proper Abel equation, those  generated by 
\begin{equation}\label{Basis2}
(-2\mu^3+3\mu x^2+x^3){\partial_x},\qquad (\mu+x){\partial_x}.
\end{equation}
These are integrable by two quadratures. For instance the case $\mu=0$ corresponds to Bernoulli equation. Finally, the algebras generated by vector fields 
\begin{equation}
(c_0+c_1x+c_2x^2+x^3){\partial_x},
\end{equation}
lead to separable differential equations, therefore integrable by one quadrature.  
\section{Canonical forms of Abel equations of the first kind}
Starting from an Abel equation of the first kind with $A_3\ne 0$, we can define a transformation 
$\bar x= x+\frac13{A_2(t)}/{A_3(t)},$ which reduces the original equation to a new one 
$\dot {\bar x}=\bar A_0(t)+\bar A_1(t)\, \bar x+\bar A_3(t)\, \bar x^3,$
where
$$
\bar A_0=A_0-\frac{A_1A_2}{3A_3}+\frac 2{27}\frac{A_2^3}{A_3^2}+\frac 13 \frac d{dt}\left(
\frac {A_2}{A_3}\right),\quad \bar A_1=A_1-\frac 12 \frac{A^2_2}{A_3},\quad\bar A_2=0,\quad  \bar A_3=A_3.
$$
In fact, a translation $\bar x=x+\beta$ transforms the given equation into:
$$\dot {\bar x}=\dot \beta+A_3({\bar x}-a)^3+A_2({\bar x}-a)^2+A_1({\bar x}-a)+A_0,$$
and then
$$
\dot {\bar x}=A_3  {\bar x}^3+(A_2-3\beta A_3) {\bar x}^2+(3\beta^2A_3-2\beta A_2+A_1) {\bar x}+\dot\beta -\beta^3A_3+
\beta^2A_2-\beta A_1+A_0.
$$
Choosing $\beta={\displaystyle \frac{A_2(t)}{3\,A_3(t)}}$ we find that $\bar A_3=A_3$, $\bar A_2=0$ and $\bar A_1$ and $\bar A_0$ 
are given by the above given expressions. When $\bar A_0=0$, we obtain a  first possibility for the   {canonical form}
\begin{equation*}
\dot {x}= A_1(t)\,  x+A_3(t)\,  x^3,
\end{equation*}
which is a Bernoulli equation for $n=3$, and therefore solvable by the change of variable $u=1/x^2$, which leads to 
the inhomogeneous linear equation 
$$\dot u+2A_1u+A_3=0\,.
$$

On the contrary, if $\bar A_0\ne 0$, under the transformation $\bar x= \bar{\bar x}\, \bar A_0,$
the equation becomes  $\bar{\bar x}'=\bar{\bar A}_3(t) \bar{\bar x}^3+\bar{\bar A}_1(x)\, y+1
$, and we obtain as a  canonical form
\begin{equation}
\dot {x}= 1+A_1(t)\,  x+A_3(t)\,  x^3.
\end{equation}
An important instance is the case $A_0=A_1=0$: the canonical forms are 
such that 
$$\bar A_1=-\frac 13 \frac{A_2^2}{A_3},\qquad \bar A_0=\frac 2{27}\frac{A_2^3}{A_3^2}+\frac 13 \frac{d}{dt}
\left(\frac {A_2}{A_3}\right).
$$

Liouville proved that if $\Phi_3$ and $\Phi_5$ are given by 
$$\begin{array}{rcl}
\Phi_3&=& A_2\,A'_3- A'_2\,A_3+3\,A_0\,A_3^2-A_1\,A_2\,A_3+\frac 29 A_2^3\\
\Phi_5&=&A_3\,\Phi'_3-3\left(A'_3+\frac 13 A_2^2-A_1\,A_3\right)  
\end{array}
$$
then   the quotient $\Phi_3^5/\Phi_5^3$ is an invariant under the structure invariance group.
 The first canonical form corresponds to $\Phi_3=0$. Any two equations with $\Phi_3=0$
 are in the same orbit, i.e. are related by the structure invariance group.
 
 For equations with the second canonical form there is a new invariant, and they are in the same orbit if and only if the
  value of the invariant is the same for both equations.
  
  There exist a two-parameter structure invariance  group for this type of equations in the case of first canonical form and  
  a one-parameter one in the cases of the second form provided that an invariant has a constant value different
  from zero.

\section{Second order Abel equation}
\label{secondorder}
 Lie's recipe
for order reduction of second-order linear differential equations leads to some Riccati equations. For instance $\ddot y=0$  
under the change of variable $y=e^{z}$ becomes ${\mathbb{D}}x=0$ with $
 {\mathbb{D}} =d/dt+ x\,,\ {\rm where}\  x=\dot z$.  Similarly $\dddot y=0$ becomes under the same
 change of variable  ${\mathbb{D}}^2x=0$. On the other hand, when comparing Abel equation with 
 Riccati equation one sees that a possible generalization to second-order equation of Abel 
 equation is  ${\mathbb{D}}_A^2x=0$, where ${\mathbb{D}}_A=d/dt + x^2$ \cite{CGR09}. The second-order Abel 
 equation (\ref{AbelEq2}) can be presented as a system of two first-order equations
$$
\left\{\begin{aligned}
 \dfrac{{dx}}{{dt}} &= v,  \cr
\dfrac{{dv}}{{d t}} &= - 4 x^2  v -  x^5,
\end{aligned}\right.
$$ 
corresponding to  the following vector field on the velocity phase space ${\mathbb{R}}^2$
$$
  \Gamma = v\,\partial_{x} + F_{A}\,\partial_{v} \,,\qquad
  F_{A} = - 4 x^2  v -  x^5.
$$
Such a nonlinear  Abel equation of second-order can be derived from the  Lagrangian 
\begin{equation}
  L_A(x,v) =  \frac{1}{(v + x^3)^2}  \,.  \label{LagAbel1}
\end{equation}
The corresponding conserved energy function is given by 
$$
 E_{L_A} = -\,\,\frac{(3 v + x^3)}{(v + x^3)^3} \,. \label{EL1}
$$
To be remarked that there exists an alternative  Lagrangian given by:
\begin{equation}
\widetilde {L}_A(x,v)   =  (3 v + x^3)^{2/3} . \label{LagAbel2}
\end{equation}
Such Lagrangians can be obtained by means of the Jacobi last multiplier theory and 
Darboux polynomials for polynomial vector fields. More specifically, Darboux polynomials for 
a polynomial vector field $X$ are polynomials ${\mathcal{D}}$ such that $X{\mathcal{D}}=f
\mathcal{D}$. The function $f$
is said to be the cofactor corresponding to such a Darboux
polynomial and the pair $(f ,\mathcal{D})$ is called a Darboux pair.

On the other side, given a vector field $X$ in an  oriented
manifold $(M,\Omega)$, a function $R$ such that $R\, i(X)\Omega$
is closed is said to be a Jacobi multiplier (JM) for $X$ \cite{NL08c}. Recall that
the divergence of the vector field $X$ (with respect to the volume
form $\Omega$) is defined by the relation
$$
 \mathcal{L}_X\Omega = ({\rm div\,}X)\, \Omega\,.
$$
Then, $R$ is a Jacobi  multiplier  if and only if $R\, X$ is a divergenceless vector field and
therefore using $$
 {\mathcal{L}}_{RX}\Omega = ({\rm div\,}RX)\, \Omega
 = [X(R)+R\, {\rm div}X]\,\Omega=0\,,
$$
 we see that $R$ is a last multiplier for $X$ if and
only if
\begin{equation}
 X(R) + R\,{\rm div}X = 0\,.       \label{Jmcond}
\end{equation}
because, for any function $f$, 
$$  X(f\, R)+f\,R\, {\rm div\,}X=(Xf)\, R+f( X(R)+R\, {\rm div\,}X).
$$

The remarkable point is that if ${\mathcal{D}}_1, \ldots,
{\mathcal{D}}_k$ are Darboux polynomials with corresponding
cofactors $f_i$, with $i=1,\ldots,k$, one can look for multiplier
factors of the form
\begin{equation}
 R=\prod_{i=1}^k{\mathcal{D}}_i^{\nu_i},    \label{Relem}
\end{equation}
and then
$$
 \frac{X(R)}{R} =\sum_{i=1}^k\nu_i\frac{X({\mathcal{D}}_i)}{{\mathcal{D}}_i}
 = \sum_{i=1}^k\nu_i\, f_i.
$$
Consequently, if the coefficients $\nu_i$ can be chosen such that
\begin{equation}
 \sum_{i=1}^k\nu_{i}\, f_i=-{\rm div\,}X,\label{expcond}
\end{equation}
we arrive to
$$
 \frac{X(R)}{R} =\sum_{i=1}^k\nu_i\,f_i=-{\rm div\,}X,
$$
what implies that $R$ is a Jacobi last multiplier for $X$.

Finally, one can prove (see e.g. \cite{NL08c}) that if $R$ is a Jacobi multiplier for a  vector field which
corresponds to a second-order differential equation,  there is an
essentially unique Lagrangian $L$ (up to addition of a gauge term)
such that $R=\partial^2L/\partial v^2$.

Actually,  Helmholtz analysed the set of conditions that a multiplier
matrix $g_{ij}(x,\dot{x})$ must satisfy in order for a given
system of second-order equations
\begin{equation*}
  \ddot{x}^j = F^j(x,\dot{x})  \,,\quad j=1,2,\dots,n,
\end{equation*}
or the corresponding system 
\begin{equation*}
\left\{\begin{aligned}
\frac{{dx^i}}{dt}&=v^i,\\
\frac{{dv^i}}{dt}&=F^i(x,v),
\end{aligned}\right.
\end{equation*}
which provides us the integral curves for the vector field 
\begin{equation}
X=v^i\partial_{x^i}+F^i(x,v)\partial_{v^i}\,,
\end{equation}
when written of the form
$$
 g_{ij}(\ddot{x}^j -F^j(x,\dot{x}))=0  \,,\quad i,j=1,2,\dots,n,
$$
to be the set of Euler--Lagrange equations for a certain Lagrangian $L$.

Each matrix solution $g_{ij}$ can be identified with the Hessian matrix of $L$,
$g_{ij} = \partial L/\partial v^i\partial v^j$, and then $L$ can be obtained by direct integration of the
$g_{ij}$ functions. 

The two first conditions just impose
regularity and symmetry of the matrix $g_{ij}$; the two other
equations introduce relations between the derivatives of $g_{ij}$
and the derivatives of the functions $F^i$. 
The
fourth set of conditions for
$g_{ij}$ is
$$
X(g_{ij}) =  g_{ik}A^k\,_{j}+ g_{jk}A^k\,_{i} \,,\quad
A^i\,_{j}=-\frac{1}{2}\partial_{v^j}{F^i} \,.
$$
When the system is one-dimensional  we have $i=j=k=1$ and then
the three first set of conditions become trivial and the fourth
one reduces to one single P.D.E.
$$
X(g) + g\,\pd{F}{v} \equiv v\,\pd{g}{x} + F
\,\pd{g}{v} + g\,\pd{F}{v}= 0 \,, \label{EqHelm}
$$
which is the equation defining the Jacobi
multipliers,
because ${\rm div}\, X=\partial_v F$.

Then, the inverse problem reduces to find the function $g$ which is a Jacobi multiplier and $L$ is
obtained by integrating  the function $g$ two times with respect to velocities. The function
$L$ so obtained  is unique up to addition of a gauge term.

Coming back to  the second-order Abel equation case, one can easily check that 
$ {\mathcal{D}}_1(x,v) = v+x^3$
is a Darboux polynomial for $ \Gamma$ with cofactor $-x^2$ since
$$
 \left(v\,\pd{}{x} + F_{A}\,\pd{}{v} \right)(v+x^3)
 = - x^2(v+x^3)\,.
$$
The  divergence of the vector field  $ \Gamma$ is $-4x^2$, and
then we see that there is a Jacobi
multiplier of the form $R= {\mathcal{D}}_1^{-4}$.
 Consequently, the Abel equation admits a
Lagrangian description by means of a function $L$ such that
$$
 \pd{^2L}{v^2} = (v+x^3)^{-4}\,,
$$
from where we obtain the Lagrangian $L=L_A$ given by (\ref{LagAbel1}).
 
 But ${\mathcal{D}}_2(x,v)=3v+x^3$ is also a Darboux polynomial for $ \Gamma$ with cofactor $-3x^2$,
 $$
 \left(v\,\pd{}{x} + F_{A}\,\pd{}{v} \right)(3v+x^3) =
 3x^2v-3(4x^2v+x^5) =-3x^2(3v+x^3)\,,
$$
and then we can find another Jacobi multiplier of the form
${\mathcal{D}}_2^{\nu_2}$ with $\nu_2=-4/3$.  The Abel equation
admits a Lagrangian description by means  of a function $L$ such
that
$$
 \pd{^2L}{v^2} = (3v+x^3)^{-4/3}\,,
$$
from where we obtain the Lagrangian $L=\widetilde{L}_A$ given by (\ref{LagAbel2}).




\end{document}